\documentclass[11pt]{article}
\usepackage{graphicx}
\usepackage{epsfig}

\begin{document}
\begin{center}
\centerline{\LARGE\bf The masses of the neutrinos}
\vspace{6mm}
\centerline{Peter S. Cooper}
\centerline{Fermi National Accelerator Laboratory}
\vspace{3mm}
\centerline{May 10, 2016}
\end{center}

\begin{abstract}
If the cosmological limits on the sum of the neutrino masses are taken
seriously we have first measurements of the masses of the neutrinos.  Using 
the Planck experiment's limit $\sum_{i=1}^3{m_i} < 230 ~meV$ and some simple 
assumptions on measurement uncertainties the mass of the heaviest neutrino 
is $63\pm11 ~meV$ and the lightest $40\pm18 ~meV$ for either hierarchy. 
\end{abstract}

A recent seminar at Fermilab reviewed the status of the sum of the masses of 
neutrinos from various cosmological measurements~\cite{Niro}.  If those limits 
are taken seriously then we already have measurements of the masses of the 
three neutrinos with some precision.  This note reduces the simple algebra of 
the neutrino masses to graphical form to illustrate this.

The cosmological limits on the sum of neutrino masses is a long and growing 
list of measurements~\cite{PDG}.  I have chosen to use the latest result of 
the Planck collaboration~\cite{Planck}, M = $\sum_{i=1}^3{m_i} < 230 ~meV$,
to work in units of meV since all the interesting numbers are cleanly 
represented to appropriate precision as integers, and to use the PDG's 
averages for the neutrino mass-squared differences:

\begin{center}
$\Delta m^{2}_{12} = m^2_{2} - m^2_{1} = 75.3 \pm 1.8 ~meV^2 $ 
$\Delta m^{2}_{32} = m^2_{3} - m^2_{2} = 2420 \pm 60 ~meV^2 ~Normal~hierarchy$
$-\Delta m^{2}_{32} = m^2_{2} - m^2_{3} = 2490 \pm 60 ~meV^2 ~Inverted~hierarchy$
$M = \sum_{i=1}^{3}m_{i} = m_1 + m_2 + m_3$
\end{center}

\begin{figure}[htbp]
\begin{center}
\includegraphics[width=\textwidth]{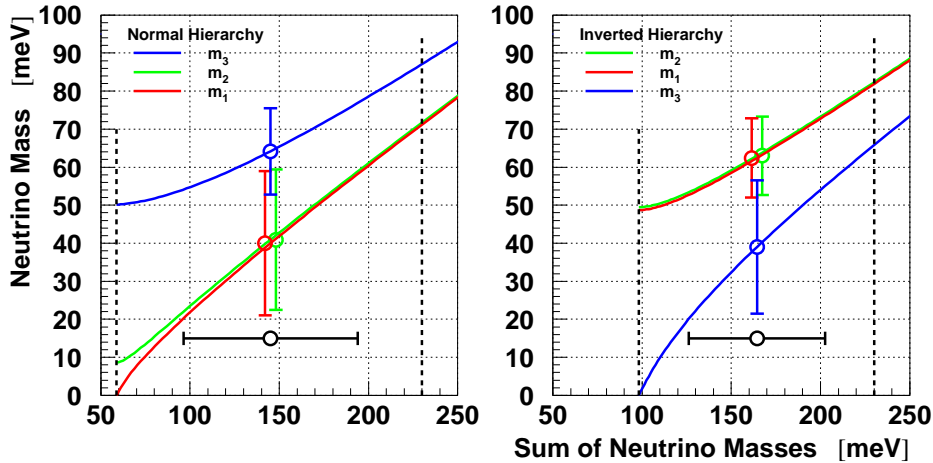}
\caption{Masses of the neutrinos as a function of the sum of the neutrino
masses.}
\label{masses}
\end{center}
\end{figure}

The dependence of the masses of the three neutrinos as a function of M, the 
sum of the masses,  are shown as the colored curves of figure~\ref{masses}.  
These curves illustrate the well known results that the minimum mass of the 
heaviest neutrino is 50~meV for either hierarchy and the minimum sum of the 
neutrino masses are 59~meV and 98~meV for the Normal and Inverted hierarchies 
respectively.  The bounds on $M$ are shown as dashed lines on 
figure~\ref{masses}: the lower bound from the non-negativity of the lightest
neutrino mass, and the upper bound from the Planck experimental limit.  

The Planck limit is a $95\% CL$ upper bound; there is a $5\%$ probability 
that the physical value is higher.  Ignoring that $5\%$ the probability 
distribution for $M$ is taken to be a box with zero probability beyond the 
limits and a constant between them representing our lack of any prior 
knowledge of the value of $M$ within the limits.  Those who have used wire 
chambers know by rote that the mean and standard deviation of a box 
distribution of width $w$ are $<w> = w/2$ and $\sigma=w/\sqrt{12}$ 
respectively.  Applying these to the normal and inverted hierarchy bounds 
yield measurements of $M = 145 \pm 49 ~meV$ and $M = 165 \pm 38 ~meV$ 
respectively.  These values are shown as the point with horizontal error bars 
on figure~\ref{masses}.  Propagating those uncertainties to the individual 
neutrino masses yields the three colored points with error bars in the figure 
which are tabulated in table~\ref{mass}. 

The short summary is that the heaviest neutrino is $63 \pm 11 ~meV$ and the 
lightest $40 \pm 18 ~meV$ independent of hierarchy with differences small 
compared to the uncertainties.  The error bars are slightly asymmetric due to 
the curvature of the mass curves, again with small differences compared
to the uncertainties.  As figure~\ref{masses} shows the three neutrino masses 
measured in this way are completely correlated: if $M$ goes down all three masses
$m_i$ go down together.

\begin{table}
\centering
\begin{tabular}{l|c|c|}  \hline\hline
 &Normal&Inverted\\
\hline
$m_1$                 & $40_{ -20}^{ +18} ~meV$  & $62_{ -10}^{ +11} ~meV$  \\
$m_2$                 & $41_{ -19}^{ +18} ~meV$  & $63_{ -10}^{ +11} ~meV$  \\
$m_3$                 & $64_{ -10}^{ +13} ~meV$  & $39_{ -19}^{ +16} ~meV$  \\
$M = \sum_{i=1}^{3}{m_i}$ &$145 \pm 49 ~meV$  &$164 \pm 38 ~meV$  \\
\hline
\end{tabular}
\caption{Computed mass values and uncertainties assuming the sum of masses
shown} 
\label{mass}
\end{table}

The Planck collaboration has published the profile likelihood  distribution 
for their previous limit of $M<260 ~meV$~\cite{Planck2}.  That result looks 
quite Gaussian ($\Delta\chi^2$ is parabolic) however $M \sim -50 \pm 155~meV$. 
They apply the Feldman-Cousins prescription to evaluate their limit based on 
the constraint that $M>0$.  This limit is physically too low, given what we 
know about neutrino oscillations, and the resulting PDF after the 
Feldman-Cousins prescription is not available.  

\begin{figure}[htbp]
\begin{center}
\includegraphics[width=\textwidth,angle=0]{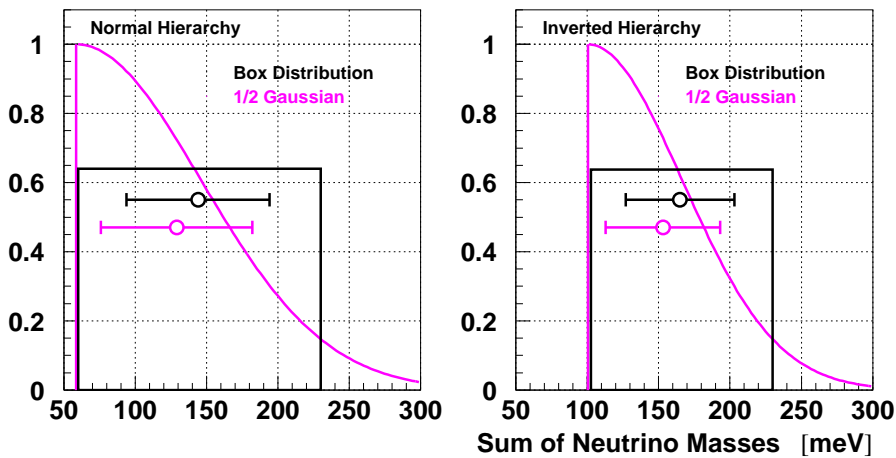}
\caption{PDFs for the sum of the neutrino masses.}
\label{pdfs}
\end{center}
\end{figure}

In order to assess the effect of the $5\%$ probability that $M > 230~meV$ a 
toy probability distribution, based on a Gaussian distribution function,
is constructed, following what Planck has reported.  A normalized 
Gaussian has two parameters: one must be fixed with the other determined by 
requiring a $5\%$ tail above $230~meV$.  Setting the mean of the Gaussian to 
the lower bound this is called a ``1/2 Gaussian'' distribution.  A more 
positive mean leads to what these distribution functions will begin to look 
like when Planck is able to make measurements rather that just set limits.  
These distribution functions are shown in figure~\ref{pdfs} with their means 
and standard deviations tabulated in table~\ref{pdf}.  The standard deviations 
for the 1/2 Gaussians increase $\sim20\%$ when evaluated out to $+\infty$ 
relative to just the bounded region.  However, those larger values are well
within $10\%$ of the standard deviation of the box distributions.  For a first 
measurement of the heaviest and lightest neutrino masses, with $20\%$ and 
$50\%$ precision respectively, uncertainties that are known to $10\%$ of 
themselves will be taken as good enough.

\begin{table}
\centering
\begin{tabular}{l|c|c|c|c|}  \hline\hline
 &limits&Normal&limits&Inverted\\
\hline
Box          &$59 - 230$   &$149 \pm 49 ~meV$&$98 - 230   $& $164 \pm 38 ~meV$\\
1/2 Gaussian &$59 - 230$   &$122 \pm 43 ~meV$&$98 - 230   $& $148 \pm 33 ~meV$\\
1/2 Gaussian &$59 - \infty$&$129 \pm 52 ~meV$&$98 - \infty$& $153 \pm 40 ~meV$\\
\hline
\end{tabular}
\caption{sum of masses mean and standard deviation for different 
distribution choices.} 
\label{pdf}
\end{table}

A most important consideration is the subjunctive with which this paper began: 
\emph{If the cosmological limits on the sum of the neutrino masses are 
taken seriously}.  If these limits get down to $<100~meV$, and are correct, 
then the Inverted hierarchy is ruled out and the neutrino masses would be 
measured with lower values and precisions of a few percent for the heaviest 
state.  These would be very important results whose reliability can, and will, 
be vigorously questioned.  The connection between the observables in 
experiments like Planck and the sum of the neutrino masses includes Physics 
modeling which is subject to systematic uncertainties.  Critical evaluation of 
those systematics are serious questions which must be left to, and defended 
by, the experts in those models and experiments.  Improvements to the 
$<100 ~meV$ level in direct measurements of a neutrino mass; the endpoint in 
Tritium beta decay, for example, will require advancements in experimental 
techniques which are, thus far, un-achieved.  
People are trying~\cite{Project8}.

I would like to thank Viviana Niro for her seminar and my Fermilab colleagues
Carl Albright, Boris Kayser, and Stephen Parke for useful conversations.


\end{document}